\documentclass[12pt,preprint]{emulateapj}

\shorttitle{Micro-Sigmoids as Progenitors of Polar Coronal Jets}
\shortauthors{Raouafi et al. 2009}

\begin{document}


\title{Micro-Sigmoids as Progenitors of Coronal Jets\\
Is Eruptive Activity Self-Similarly Multi-Scaled?}


\author{N.-E. Raouafi\altaffilmark{1}, M. K. Georgoulis\altaffilmark{2}, D. M. Rust\altaffilmark{1}, \and P. N. Bernasconi\altaffilmark{1}}
\affil{\altaffilmark{1}The John Hopkins University Applied Physics Laboratory, Laurel, MD, USA}
\affil{\altaffilmark{2}Research Center for Astronomy and Applied Mathematics, Academy of Athens, Athens GR-11527, Greece}
\email{Nour.Eddine.Raouafi@jhuapl.edu}




\begin{abstract}
Observations from the X-ray telescope (XRT) on Hinode are used to study the nature of X-ray bright points, sources of coronal jets. Several jet events in the coronal holes are found to erupt from small-scale, S-shaped bright regions. This finding suggests that coronal micro-sigmoids may well be progenitors of coronal jets. Moreover, the presence of these structures may explain numerous observed characteristics of jets such as helical structures, apparent transverse motions, and shapes. In analogy to large-scale sigmoids giving rise to coronal mass ejections (CMEs), a promising future task would perhaps be to  investigate whether solar eruptive activity, from coronal jets to CMEs, is self-similar in terms of properties and instability mechanisms.
\end{abstract}

\keywords{Sun: corona --- Sun: activity --- Sun: X-rays, gamma rays --- Sun: UV radiation --- Sun: magnetic fields}
 
\section{Introduction}

High-resolution data from recent solar missions (e.g., Yohkoh, SOHO, STEREO, and Hinode) reveal the intense dynamical activity in the solar atmosphere, in particular, in the lower solar corona. This is illustrated by the multi-scale, ubiquitous activity (i.e., flares, spicules, surges, and jets) that may be evidence of magnetic reconnection events (see Shibata et al. 2007 and references therein).

Observations from the Soft X-ray Telescope (SXT; Tsuneta et al. 1991) on Yohkoh (Ogawara et al. 1991) led to the discovery of the bright, transitory, and sharply-collimated X-ray jets (Shibata et al. 1992; Strong et al. 1992) that are usually associated with energy release in the underlying X-ray-bright points (XBPs), emerging flux regions (EFRs), or active regions (ARs). Radio observations provided evidence for non-thermal processes occurring at the sites of coronal jets (e.g., type III radio bursts; see Kundu et al. 1994). However, the spatial and temporal resolution of imaging instruments prior to the Hinode mission (Kosugi et al. 2007) was too limited to fully resolve these coronal features and the fine structure of their sources (e.g., XBPs), or to follow their temporal evolution in detail, especially in case they are triggered in coronal holes.

The X-ray telescope (XRT; Golub et al. 2007) on Hinode provides data with unprecedented spatial resolution and cadence, which allow one to resolve the sources and evolution of small-scale solar structures (e.g., XBPs) leading to the eruption of coronal jets. The latter occur almost everywhere in the solar corona (see Shibata et al. 1992), but they are most prominent in the coronal holes (i.e., open field regions) due to the low background emission. Jets are characterized by their collimated appearance (typical length and width of $10^4-10^6$~km and $\sim10^4$~km, respectively; see Cirtain et al. 2007) and high-temperature emission in the X-ray and EUV wavelengths. Coronal jets often occur in conjunction with other phenomena that are also likely due to magnetic reconnection, such as H$\alpha$ surges (Rust 1968; Canfield et al. 1996) and polar plumes (Raouafi et al. 2008 and Raouafi 2009).

The distinctive collimated structure of coronal jets inspired the ``anemone model,'' in which a simple dipolar magnetic structure is embedded into a background of open fields. The anemone-shaped structures are widely believed to form through the reconnection of emerging dipolar, magnetic loop systems with the open -background- coronal field. Part of the trapped plasma within the emerging flux system is released in a narrow, collimated, presumably open structure, as suggested by SXT and XRT observations. Mixed polarity magnetic fields are observed at the base of jets (Canfield et al. 1996; Wang 1998). Simulations using the anemone model have been achieved in 2D (Yokoyama \& Shibata 1995), 2.5D (Karpen et al. 1998; Archontis et al. 2007), and 3D (Archontis et al. 2005 \& 2006) numerical simulations. Per this model, the jet may appear in different shapes depending on the reconnection site: jets appear either with an inverse-Y- or $\lambda$-shape depending on whether the reconnection occurs at the cusp or at the legs of the emerging dipolar loops, respectively. However, since the anemone model involves interchange reconnection between closed and open field lines, the resulting jet has a single thread. Multi-threaded jets cannot be explained by the simple anemone model.

\begin{figure*}[!t]
\epsscale{1}
\plotone{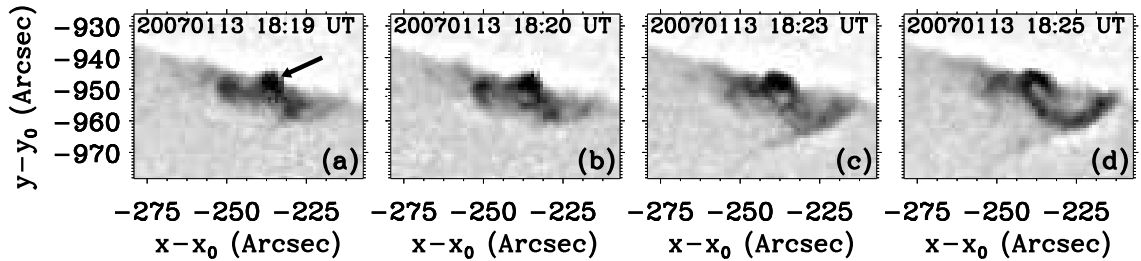}
\caption{The fine structure of an XBP at the solar limb in the southern polar coronal hole as observed by XRT (Al\_poly filter). The morphology of the different loops suggests a feature more complex than a dipolar loop system. The arrow points to the micro-flare location that caused the expansion of the loop system (panel b) followed shortly by the eruption of the two-thread jet (downward in panels c-d).\label{fig1}}
\end{figure*}

On the other hand, complex coronal X-ray jets showing an apparent untwisting motion have been reported by several studies using SXT observations (see Shimojo et al. 1996; Canfield et al. 1996). Recently, Patsourakos et al. (2008) used STEREO/SECCHI/EUVI (Howard et al. 2008) images to determine the 3D topology of a jet event and found evidence for unwinding magnetic fields suggesting that the jet may be related to helical field lines. Nistic\`o et al. (2009) carried out a statistical study of coronal jets using XRT observations of 79 events. They found that among 61 characterized events 31 ($\sim50$\%) show untwisting behavior.

Two distinct approaches have been attempted to interpret the observed untwisting in jets. The first assumes that the emerging dipole tube is twisted, allowing for helicity transfer to the newly-formed, open jet following magnetic reconnection (see Canfield et al. 1996). Moreno-Insertis et al. (2008) developed 3D magnetohydrodynamical (MHD) simulations based on the interaction of an emerging twisted flux tube with a pre-existing, open magnetic field. The jet occurs on one side of the flux tube where the tube and background-open fields are counter-aligned. Although the simulations qualitatively reproduce the apparent properties of the observed jets, it is not clear how this model accounts for the helical structure often observed in jets (e.g., Wang et al. 1998; Jiang et al. 2007). The second approach consists of artificially rotating the base of the bright point allowing for a build-up of twist, as simulated by Pariat et al. (2009). The physical mechanism responsible for the twisting motions remains, however, unclear.

Rust \& Kumar (1996) observed transient S-shaped brightenings (called sigmoids) prior to the eruption of coronal mass ejections (CMEs). They suggested that sigmoids have a rather complex helical flux-rope structure. Upon eruption, commonly attributed to the helical kink instability, a flux rope expands forming the CME while the remaining structure is either a loop arcade perpendicular to the spine of the pre-eruption sigmoid or a diffuse cloud (Rust \& Kumar 1996). Coronal sigmoids can be both transient and long lasting. Canfield et al. (1999) analyzed 117 active regions observed by SXT and found that, regardless of size, ARs with sigmoidal X-ray morphologies are 68\% more likely to be eruptive than non-sigmoidal ARs. Canfield et al. (2007) further reported that $\sim83$\% of the eruptions associated with sigmoids are detected as CMEs.

To shed light on the nature of the source regions of coronal jets, we took advantage of the exceptional data quality from XRT, EIS, and EUVI. We studied the morphology and fine structure of XBPs leading to coronal jets in conjunction with their characteristics (i.e., untwisting motions, transverse apparent motions, shapes, etc.). We consider scenarios alternative to the anemone model that may account for the characteristics of jets and their morphology (i.e., open or closed structures). Ultimately, we address the relationship between polar coronal micro-sigmoids and X-ray jets. Interestingly, we were unable to find published studies linking these two manifestations of coronal activity.

\section{Observations and Data Analysis}

\begin{figure*}[!t]
\epsscale{1}
\plotone{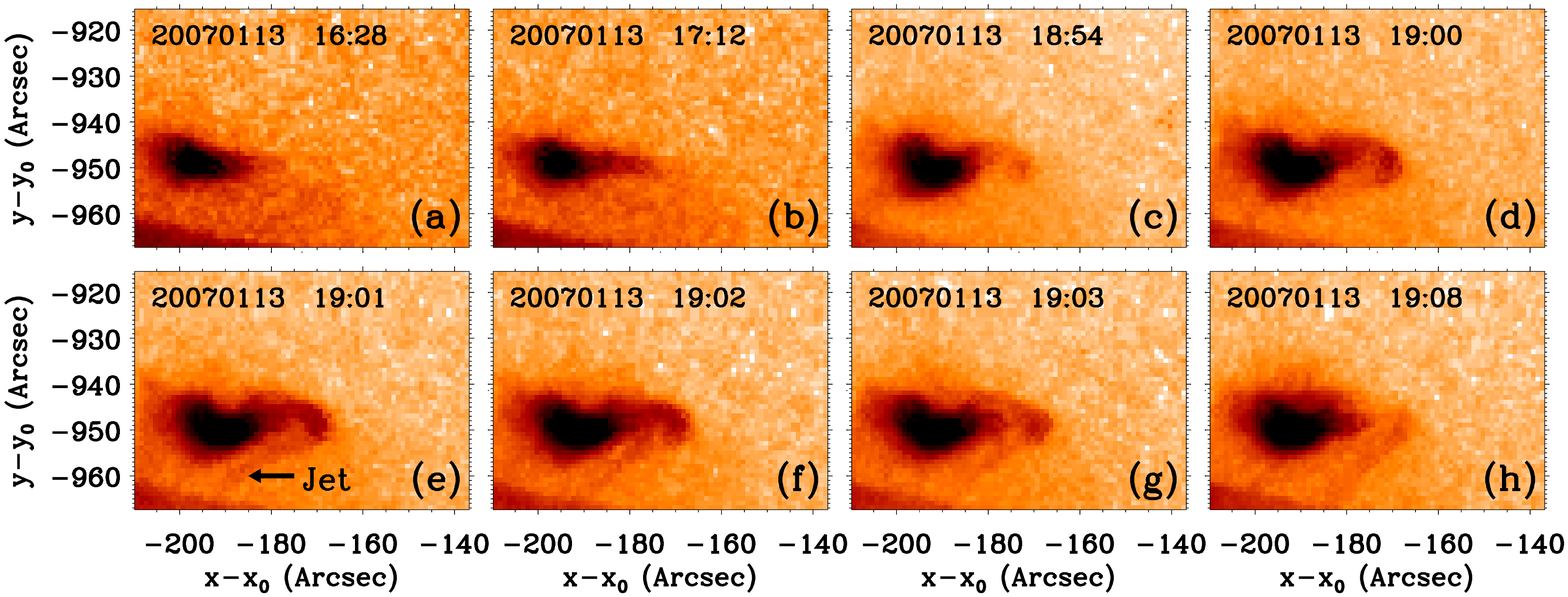}
\plotone{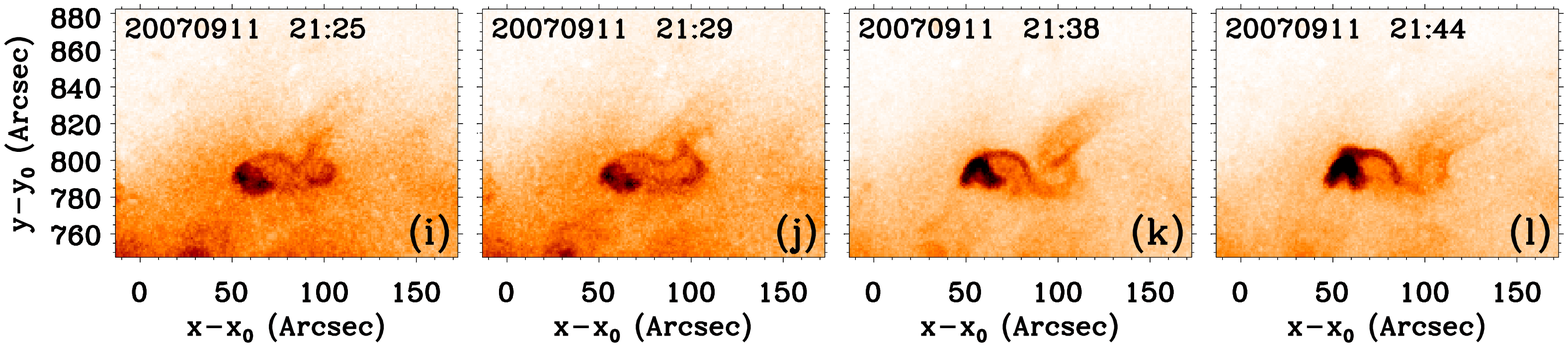}
\caption{XRT observations with Al\_poly filter of the evolution of XBPs. (a-h): XBP in the southern polar coronal hole developing into a micro-sigmoid that yielded a jet ({\bf{pointed}} by the arrow in panel e). The jet appeared at different locations along the sigmoid. (i-l): Sigmoidal structure erupting into a prominent jet in the northern polar region. Notice the two simultaneous sigmoids in panel (j). Dates and times of the observations are plotted in each panel. \label{fig2}}
\end{figure*}

Images recorded by XRT have unmatched spatial and temporal resolution ($\sim1\arcsec$ and $<1$~minute, respectively). Furthermore, the different filter systems allow a temperature coverage from $\sim1$~MK up to $\sim20$~MK. XRT observations of the solar coronal holes are our primary data source to delineate the origins, formation, and evolution of X-ray jets. The data are corrected for instrumental effects utilizing XRT calibration procedures available on SolarSoft (sohowww.nascom.nasa.gov/solarsoft). Data from the Extreme UV Imaging Spectrometer (EIS; Culhane et al. 2007) on Hinode and EUVI were also utilized to obtain complementary information on some events.

Fig.~\ref{fig1} displays XRT images of a bright point at the solar limb within the southern polar coronal hole. The XBP erupted into a jet at 18:23~UT on January 13, 2007. The spatial resolution is sufficient to resolve, at least partially, the XBP's complex fine structure. The complexity of the loops suggests that the XBP's magnetic fields are also complex. A small brightening - a microflare - occurred at the base center of the loop structure at about 18:19 UT (arrow on Fig.~\ref{fig1}a), and it likely triggered the jet. Figs.~\ref{fig1}a-c show the loop system expanding up to the instant of the jet's eruption. The jet is double threaded (the thread on the left appears shorter and dimmer than the one on the right; see Fig.~\ref{fig1}d). This and other cases motivated us to study the fine structure of XBPs and their evolution prior to release of a coronal jet. We found that bright points may be more complex than previously thought (i.e., dipolar structure), and jets may be the result of more sophisticated physical mechanisms than magnetic reconnection between a simple dipolar loop system and open background fields.

Fig.~\ref{fig2} displays the different stages of evolution of two (initially simple) XBPs located in the southern (Figs.~\ref{fig2}a-h) and northern (Figs.~\ref{fig2}i-l) polar coronal holes. Both XBPs evolved into more complex structures that led to jet eruptions. The initial appearance of the bright region of January 13, 2007, is simple and may suggest a system of dipolar loops. The brightness of the XBP increased with time and its shape changed dramatically. Fig.~\ref{fig2}b shows a section of the XBP extending westward. The newly appearing section is dimmer than the initial bright region but clearly noticeable. The detailed structure of the XBP at its peak brightness is more complex and significantly different from its initial appearance. Figs.~\ref{fig2}c-g displays an S-shaped structure suggesting a coronal micro-sigmoid that is oriented roughly in the solar East-West direction. The size of the micro-sigmoid is about 30{\arcsec} by 10{\arcsec}, which is approximately {\it{one tenth}} the size the large-scale sigmoid studied by McKenzie \& Canfield (2008). Like most large-scale, active-region and quiet-sun alike, sigmoids, this micro-sigmoid is not uniformly bright.

A jet erupted first on the western end of the sigmoid (Fig.~\ref{fig2}e). The jet appeared at slightly different locations along the sigmoidal structure (see Figs.~\ref{fig2}e-h). This may reflect a filamentary structure showing that the jet is probably composed of different strands. The sequence of heating and cooling of different strands probably dictate the appearance and disappearance of the different threads. It is also likely that the different locations correspond to multiple magnetic reconnection sites at different times. In the aftermath of the jet eruption, the micro-sigmoid section where the jet erupted was altered and the brighter section became more diffuse (Fig.~\ref{fig2}h). In the course of the jet appearance, the XBP was progressively loosing its sigmoidal shape.

\begin{figure*}[!t]
\epsscale{1}
\plotone{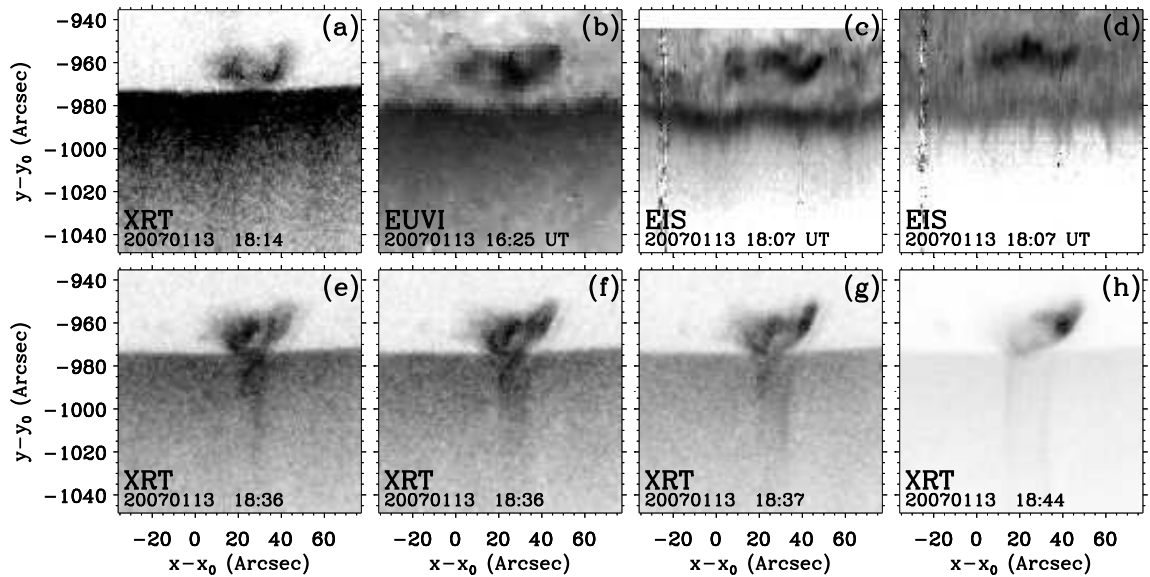}
\caption{Top: (a) XRT/Al\_poly; (b) STEREO/SECCHI/EUVI 195~{\AA}; (c-d) EIS 195~{\AA} and 257~{\AA}. The XBP suggests the presence a polar micro-sigmoid. Bottom (e-h): evolution of a helical jet as observed by XRT/Al\_poly. \label{fig3}}
\end{figure*}

The event of September 11, 2007, is quite intriguing. The XBP showed different sharply defined, sigmoidal structures as it evolved (Figs.~\ref{fig2}i-l). This suggests that each sigmoid contributes to the jet and disappears before the appearance of the following one. This structure is approximately 55{\arcsec} long and 20{\arcsec} wide ($\sim20$\% the size of McKenzie \& Canfield's event). Remarkably, at some point two different sigmoids, with spines roughly parallel to each other, co-exist over a small spatial region (Fig.~\ref{fig2}j). A jet erupted first at the center of the XBP (Fig.~\ref{fig2}i) and moved westward end (Fig.~\ref{fig2}k-l) apparently forcing the entire structure to loose its sigmoidal shape.

Fig.~\ref{fig3} shows another interesting event that occurred close to the southern solar pole on January 13, 2007. The jet took place above a relatively diffuse bright region. At first glance, it would appear that the jet might well fit the anemone jet model. However, although the background scatter around the XBP is relatively high, XRT images indicate that the XBP structure is relatively complex and composed of different extended sections (Fig.~\ref{fig3}a). However, X-ray data do not provide definitive evidence regarding the structural nature of the XBP. EUVI images (\ion{Fe}{12}~195~{\AA}) recorded around 16:24~UT ($\sim2$ hours prior to the jet eruption) indicate also the presence of a seemingly sigmoidal bright region (Fig.~\ref{fig3}b). About half an hour later, the EUV structure became increasingly diffuse, eventually losing its S-shape, which is probably due to its heating or cooling sequence. More solid evidence for the sigmoidal nature of the XBP was provided by EIS. Raster sequences recorded at \ion{Fe}{12}~195~{\AA} and \ion{Fe}{12}~257~{\AA} at about 18:06~UT clearly show the micro-sigmoid (Figs.~\ref{fig3}c-d). The latter is also oriented in the East-West direction and is the source of a prominent jet that erupted at 18:10~UT (Figs.~\ref{fig2}e-h).

A prominent, initially inverse-Y jet erupted from the same region at about 18:31~UT. The interesting aspect of this event is that it showed unwinding behavior (see Figs.~\ref{fig3}e-h), which is evidence of helicity transfer from closed to open  field lines. The unwinding threads evolved into a simpler, double-threaded structure. The right-hand thread dimmed faster than the other one leaving an apparent $\lambda$-shaped jet (Figs.~\ref{fig3}h). The present case shows the source of the helical magnetic structure of the jet: eruptive coronal sigmoids have helical magnetic fields, regardless of their nature (i.e., flux rope or a sheared arcade). Following the eruption, helicity is transported and redistributed over the entire post-eruption flux rope, as in case of a large-scale erupting filament with a subsequent CME. The double threaded structure of the present event may also suggest a complete eruption of the micro-sigmoid, perhaps suggesting a``micro-CME''. In this case, the jet is at the feet of a magnetically closed structure, nearly like the standard model of a large-scale post-eruption flux rope.

\begin{figure*}[!t]
\epsscale{1}
\plotone{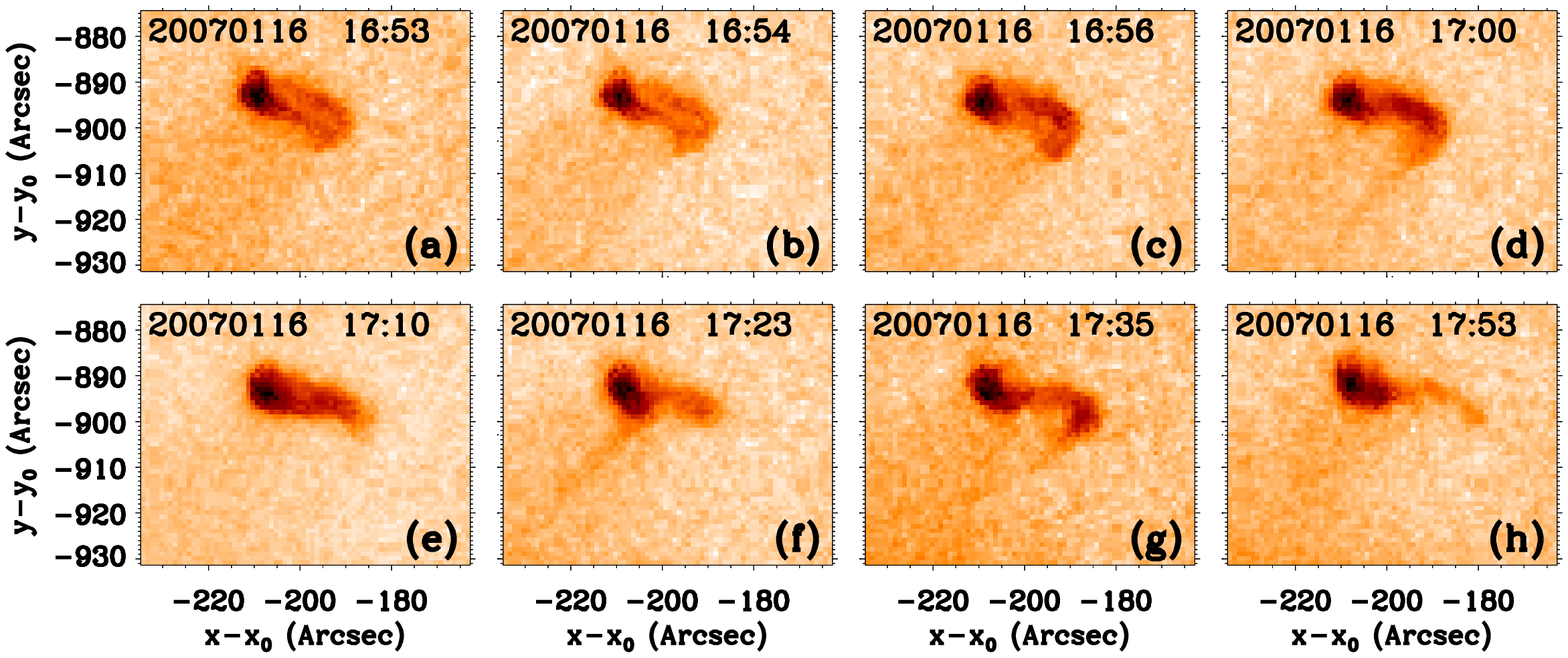}
\plotone{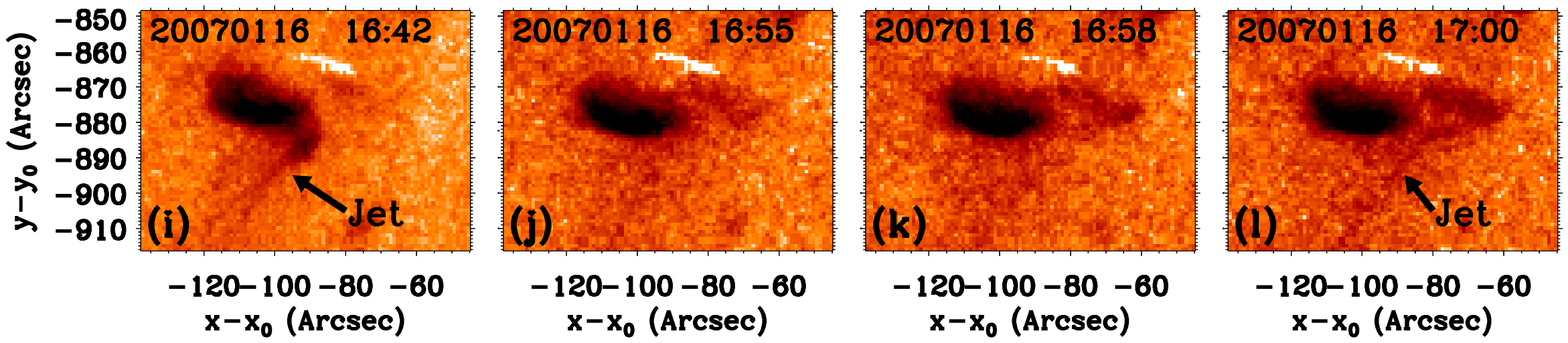}
\caption{Same XRT filter as in Fig.~\ref{fig2}. The XBP in panel (i) suggests a NE-SW oriented micro-sigmoid. However, the real structure is larger and oriented East-West as shown in panels (j-l). (a-h): Evolution of J-shaped loop systems into a sigmoid that yielded multiple, non-localized jets. (i-l): True sigmoidal nature of an apparently simple loop system. The dim-section is only seen after extensive contrast . \label{fig4}}
\end{figure*}

Fig.~\ref{fig4} shows two more examples of polar micro-sigmoids. The XBP shown in Figs.~\ref{fig4}a-b has J-shaped loop structures that merged into one brighter sigmoid where different jets erupted at different times and from different locations along the sigmoid. This evolution is similar to the large scale case studied by Mckenzie \& Canfield (2008). The jet is apparently moving transverse to its outflow direction (see Fig.~\ref{fig4}c-g). Notice also the substantial morphological change of the source regions prior to and after the jets (Fig.~\ref{fig4}h).

The jet event in Fig.~\ref{fig4}i (16:42~UT) emanates from the right-end of the bright feature yielding a $\lambda$-shaped jet. The bright region extended westward, however, the structure's newly appearing, dim section is hardly noticeable. Extensive contrast enhancement reveals that the whole structure is S-shaped, more extended, and East-West oriented (Fig.~\ref{fig4}j-l). Another jet erupted at 17:00 UT from the dim section of the sigmoid giving rise to a $\lambda$-shaped feature.

The events shown in Figs.~\ref{fig3}-\ref{fig4} appear in different shapes at different times: inverse-Y jets in Fig.~\ref{fig3} and $\lambda$-jets in Figs.~\ref{fig2},\ref{fig4}. Apparent transverse motions of the jets are also inferred in Figs.~\ref{fig2},\ref{fig4}. Assuming that the jets are the result of micro-sigmoid eruptions, the inverse-Y jets may correspond to complete eruption of the sigmoid as in Fig.~\ref{fig3}. The $\lambda$-jets, on the other hand, may be the result of partial eruption at one end of the sigmoidal structure such as the event in Fig.~\ref{fig2}. The apparent transverse motion of jets may also correspond to independent but successive eruptions of different loop strands within the sigmoid. Assuming that the S-structure is a conglomeration of loops (e.g., a sheared arcade), if the reconnection starts at one end of the arcade and proceeds toward the other end, a series of jets appear like one jet moving transverse to its outflow direction. Fig.~\ref{fig6} displays a cartoon model for transversally moving jets. It is based on the model by Titov \& D{\'e}moulin (1999). The loop systems related to the separatrix surfaces associated with bifurcated bald patches at each end of the flux rope may reconnect with the background coronal magnetic field. Transversally moving (non-twisting) jets may be the result of successive magnetic reconnections of the separatrix-surfaces' loop with the open -background- fields. Untwisting jet may correspond to eruption of the entire flux rope.

Table~\ref{table1} lists a set of micro-sigmoids in the north (NP) and south (SP) polar coronal holes. Although the analyzed data sets are sparse, a few tens of these events were identified. Most cases resulted into coronal jet eruptions. This shows that such manifestations occur frequently in the coronal holes. It is necessary to analyze larger data sets in order to characterize coronal micro-sigmoids and their relation with jets. Moreover, instruments with significantly higher spatial and temporal resolutions than the present ones may reveal more of these structures.

\section{Discussion}

High-resolution images recorded by XRT provide evidence for coronal X-ray jets emanating from small-scale (micro-) sigmoids at the polar coronal holes. The resolved structure of some bright regions is more complex than previously assumed. The new data allowed us to follow in detail the evolution of these regions prior to polar jet eruptions. Our findings are important for understanding the nature of the physical processes that drive coronal jets and related features, such as polar plumes, H$\alpha$ surges, and spicules. Several characteristics of jets (e.g., untwisting, transverse motions, and shapes) can now be explained.

Since the observation of coronal X-ray jets by SXT, the data have been getting steadily better in terms of spatial resolution and temporal cadence. Although the spatial resolution of XRT is not sufficient to resolve very small BPs, the XRT data make it possible to infer useful information about these features and the associated small-scale activity (jets, plumes, surges, spicules). Inadequate resolution previously led to the assumption that BP structure is a typical magnetic dipole and that the anemone model of jets provides an adequate description of BP activity. It has been assumed that the emerging dipolar magnetic field reconnects with the open coronal background field and that the previously trapped plasma is channeled into a collimated beam on open magnetic field lines. Although the anemone model has been shown to reproduce many morphological features of coronal jets, it also exhibits shortcomings in explaining other properties, such as helical structures (Patsourakos et al. 2008) and apparent transverse motions (see Savcheva et al 2009) of numerous jet events.

\begin{figure*}[!t]
\begin{center}
\includegraphics[scale=.3,angle=-90]{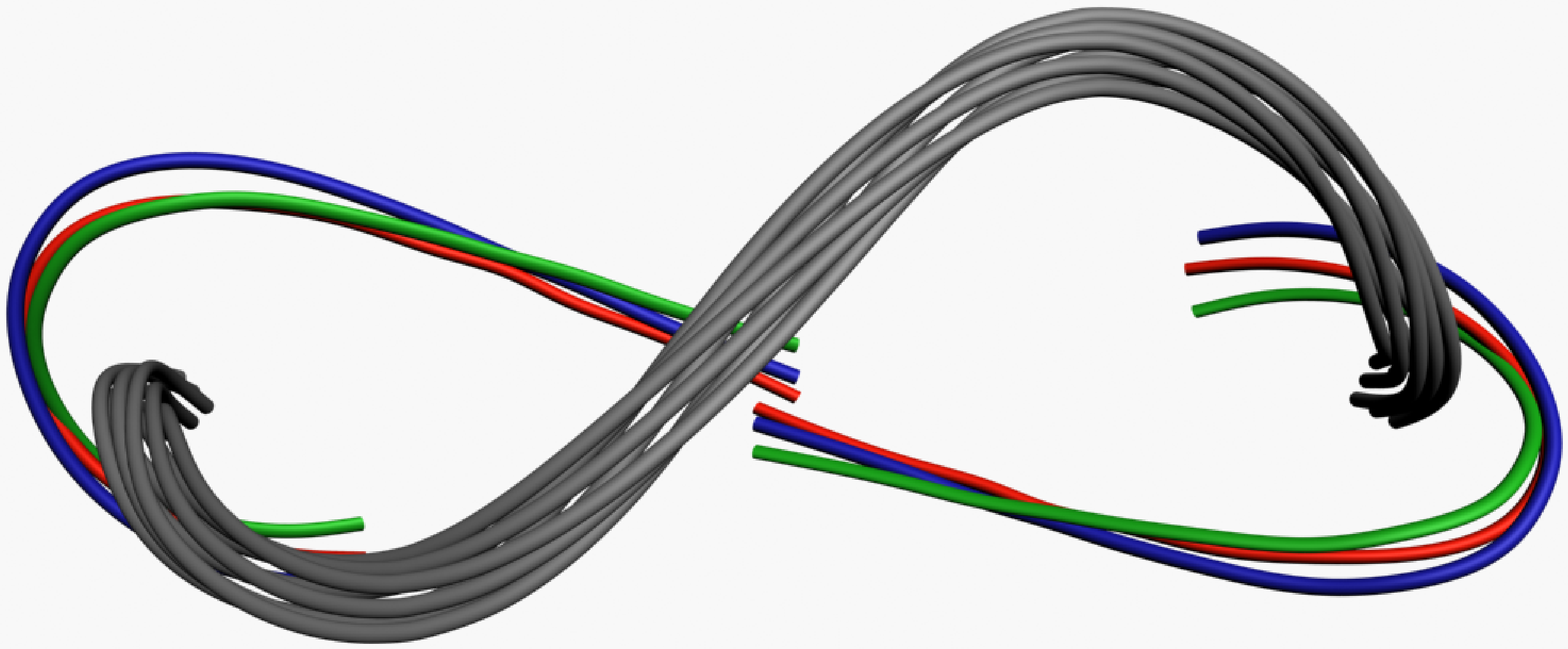}
\includegraphics[scale=.3,angle=-90]{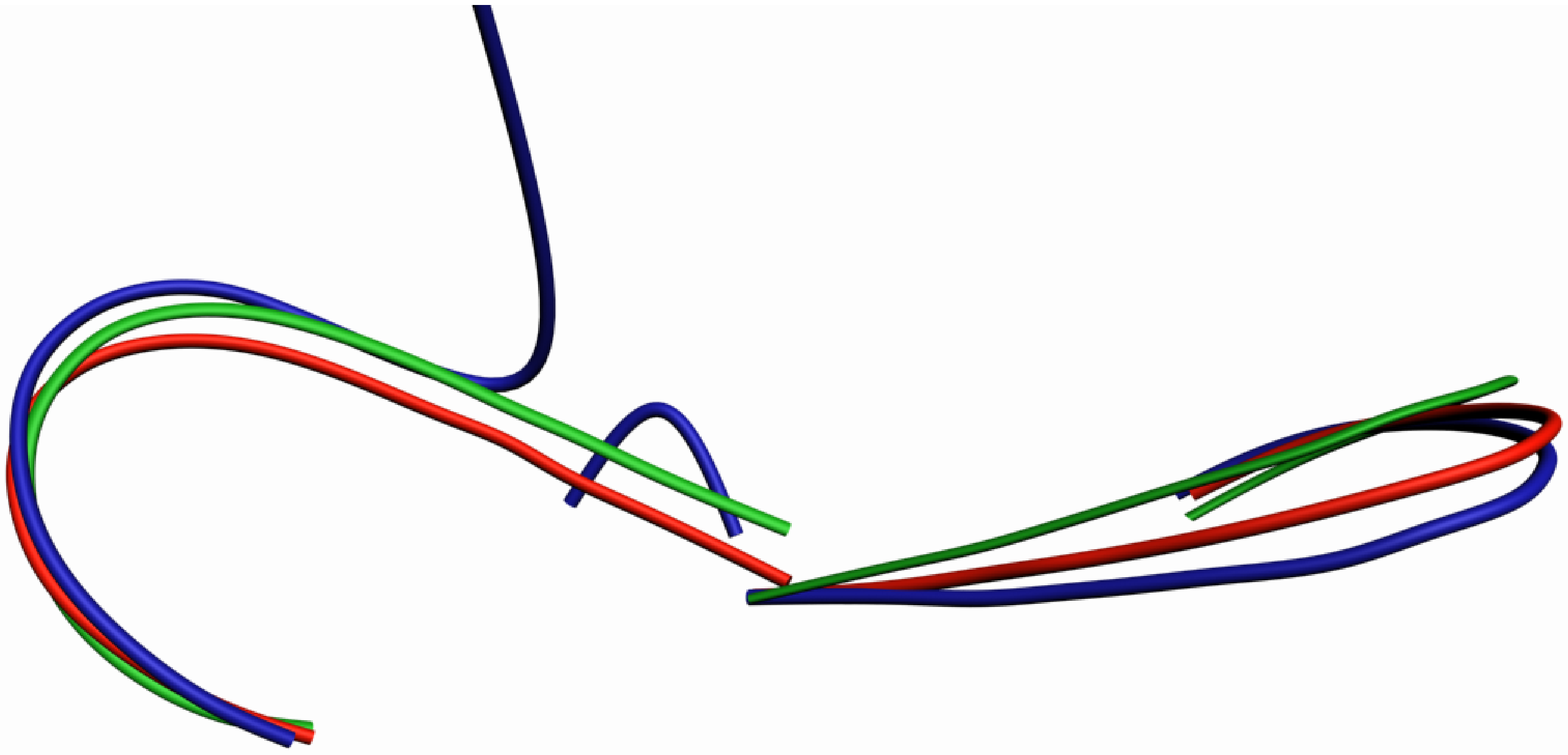}
\includegraphics[scale=.3,angle=-90]{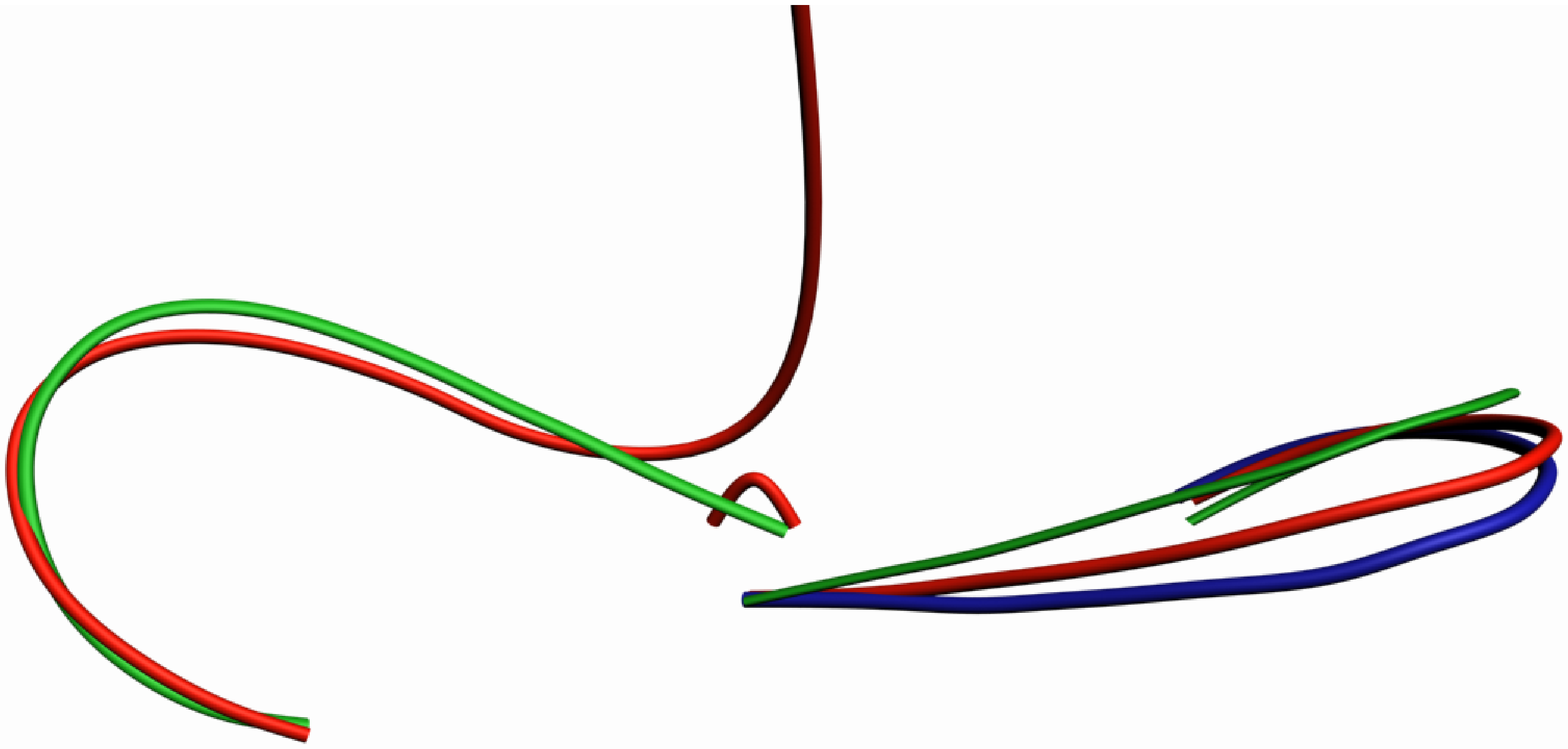}
\includegraphics[scale=.3,angle=-90]{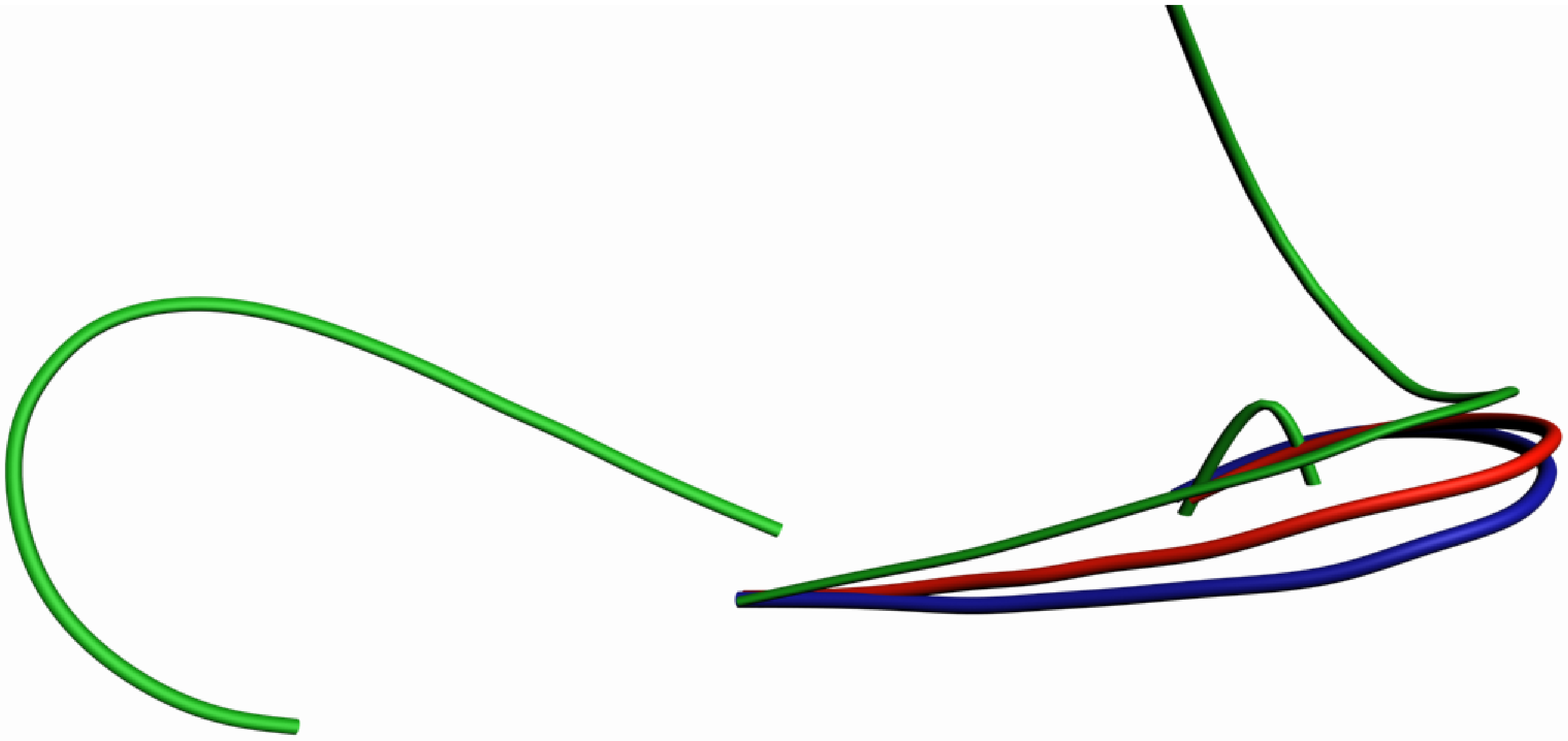}
\caption{Cartoon model for transversally moving jets. The entire structure is composed of a flux rope (grey feature in the leftmost panel) and {\bf{a bundle of loops}} that form separatrix surfaces associated with bifurcated bald patches at each end of the flux rope according to the model by Titov \& Demoulin (1999). The flux rope is removed from the other three panels for clarity sake. These panels represent a time series of successive magnetic reconnections along the structure that result in an apparently moving jet (time is running top to bottom, left to right). The loop colors are only for clarity. \label{fig6}}
\end{center}
\end{figure*}

The outstanding observations of XRT, EIS, and EUVI provide unprecedented insight into the nature of the source regions of coronal jets. The presence of micro-sigmoids at the base of jets can explain several of their properties:
\begin{itemize}

\item {\bf{Jet helical structure:}} Nistic\`o et al. (2009) carried out a statistical study on a sample of 79 jets. They found that among the classified events (i.e., 61) more than 50\% (i.e., 31 events) showed helical patterns. This shows that helical structures are quite common in coronal jets. It is also well established (from large-scale solar eruptive activity) that the eruption of coronal sigmoids, regardless of their detailed structure (flux rope [Rust \& Kumar 1996] or sheared arcade [Pneuman 1983]), leads to helical structures. Our observations of jets emanating from polar micro-sigmoids provide a natural explanation of the observed unwinding behavior of these structures. This is unlike numerical simulations based on the anemone model where the source of helicity is introduced heuristically and lacks observational support.

\item {\bf{Jet transverse motions:}} Savcheva et al. (2009) carried out a statistical study of polar coronal jets focusing on the transverse (i.e., perpendicular to the jet axis) motions of jets. They found that the direction of these motions of non-twisting jets depends upon the field polarity, which they interpreted to mean that flux emergence in the polar regions may follow a Hale-like law. These results are outstanding; however, the source of the transverse motion itself remains unclear. The fact that coronal sigmoids are formed by several loop systems that are arranged to collectively produce the observed S-shape may also provide an explanation of the reported lateral motions of jets. Fig.~\ref{fig6} shows an example model of successive reconnections of different loops along the sigmoidal flux rope that provide sequential small jet events which collectively result in an apparently moving structure from one side of the structure to the other side. This behavior is also shown by some of the events presented here (see Figs.~\ref{fig2},\ref{fig4}).

Vorpahl (1976) studied detailed evolution of a flare at 18:27~UT on September 5, 1973. She found evidence for sequential brightenings of different X-ray features along the axis of the arcade magnetic field where the flare occurred, suggesting the propagation of a triggering disturbance from one side to the other in the flare core. A magnetosonic wave was proposed to explain the observations. Vorpahl's moving X-ray features may be the large-scale equivalent of the apparently moving jets. The apparent motion may correspond to the successive eruptions of different magnetic features due to the propagation of some triggering instability within the pre-flare structure.

\item {\bf{Micro-CMEs:}} Some other events present a CME-like (micro-CME) structure. It is well known that most sigmoid eruptions are accompanied by CMEs. The observed micro-sigmoids in the polar coronal holes may help to understand CMEs.

\item {\bf{Jet shapes:}} The complete and partial eruption of micro-sigmoids into jets provides an explanation of the different shapes of coronal jets. The anemone model suggests that the inverse-Y- and $\lambda$-shaped jets are the result of the reconnection between the emerging dipolar magnetic field and the open background coronal field, where the reconnection takes place at the cusp and base of the emerging flux system, respectively. However, Figs.~\ref{fig2},\ref{fig4} suggest that $\lambda$-shaped jets are the result of reconnection at one end of the sigmoidal structure. This is particularly evident when the non-reconnecting end of the sigmoid is brighter. On the other hand, Fig.~\ref{fig3} shows an inverse-Y jet corresponding to the eruption of the entire micro-sigmoid or a section, but not just the edge, of the sigmoid.

\end{itemize}

Some properties of large-scale sigmoids are also found at small scales (e.g., J-shaped loops that merge into a single erupting sigmoid as found by McKenzie \& Canfield 2008). The eruption of micro-sigmoids, in case the overlying solar magnetic field allows, naturally leads to injection of magnetic helicity into the heliosphere. Therefore, while CMEs inject large amounts of helicity per event, micro-CMEs may be much more frequent providers of heliospheric helicity that, however, involves much smaller amounts of helicity in each eruption than typical CMEs. Such ubiquitous small-scale helicity injections have been reported very recently using wavelet-enhanced coronagraph observations (Rust et al. 2009) and may explain why there is a delay in the observed travel times of energetic particles accelerated in micro-flare sites (see Rust et al. 2008). Indeed, if heliospheric field lines are helical the particle path length from Sun to Earth is greater than 1.2 AU of the idealized Parker's spiral, thus causing the observed delay.

The findings reported here may stimulate efforts for a more fundamental understanding of solar magnetism and subsequent eruptive activity. Non-ideal instabilities leading to solar eruptions are fueled by localized magnetic energy dissipation that is known to be a small-scale process, constrained by the localized nature of magnetic reconnection (e.g., Priest \& Forbes 2000, Parker 2004). In our mechanism (Fig. 6) coronal jets are indeed due to ubiquitous magnetic reconnection, in line with Shibata et al. (2007), among others. But if the reconnected magnetic fields are helical, then helicity is proportional to the square of the magnetic flux enclosed by the sigmoidal, helical flux rope (e.g., Moffatt \& Ricca 1992). Contrary to magnetic energy release, magnetic helicity is injected from smaller to larger spatial scales, thus exhibiting an inverse cascade (Einaudi et al. 1996), and is roughly conserved during magnetic reconnection (e.g., Berger 1999). Therefore, the helicity injected to larger scales via the disappearance of micro-sigmoids, possibly giving rise to micro-CMEs, is much smaller than the helicity of typical CMEs given the much smaller reconnected magnetic flux. Whether elementary Òmicro-injectionsÓ of helicity relate to the large helicity injections of CMEs in a scale-invariant, self-similar manner is a very interesting point that is yet to be clarified. Albeit still unclear, it would be wise not to rule out this scenario given the ubiquitous manifestations of self-similarity in the solar atmosphere, ranging from active regions (Harvey \& Zwaan 1993) to sizes of hard X-ray flares (Crosby et al. 1993), to microflares and ÒnanoflaresÓ (Aschwanden \& Parnell 2002; Christe et al. 2008), to Ellerman bombs (Georgoulis et al. 2002), to even CME masses and kinetic energies (Vourlidas et al. 2002; Vourlidas \& Patsourakos 2004), among other works. Self-similarity may be viewed as the result of the turbulent evolution in the solar atmosphere (e.g., Georgoulis 2005). It would indeed be remarkable if future analyses showed that the reported power laws in CME size can be extended to smaller scales to adequately describe the statistical properties of coronal jets, thus ÒupdatingÓ the physics of these small-scale instabilities to that of micro-CME phenomena.

\acknowledgments

Hinode is a Japanese mission developed and launched by ISAS/JAXA, with NAOJ as a domestic partner and NASA and STFC (UK) as international partners. It is operated by these agencies in cooperation with ESA and NSC (Norway). The STEREO/SECCHI data used here are produced by an international consortium of the NRL (USA), LMSAL (USA), NASA GSFC (USA), RAL (UK), Univ. Birmingham (UK),MPS (Germany), CSL (Belgium), IOTA(France), and IAS (France). N.-E. Raouafi's work is supported by NASA grant \# NNX08AJ10G.

\clearpage






\clearpage

\begin{deluxetable}{ccccccl}
\tabletypesize{\scriptsize}
\tablecaption{Sample of eruptive and non-eruptive polar micro-sigmoids and their relation to coronal jets.
The times (in UT) correspond to the appearance and disappearance of the sigmoid and the jet, respectively. \label{table1}}
\tablewidth{0pt}
\tablehead{
\colhead{Location} & \colhead{Date} & \multicolumn{2}{c}{Sigmoid} &  \multicolumn{2}{c}{Jet} & \colhead{Comment}
}
\startdata

SP & 20070113 & 18:14 & \nodata & 18:33 & 18:58 & \\
SP & 20070113 & 18:50 & 19:10 & 19:01 & 19:15 & \\
SP & 20070113 & 19:17 & \nodata & \nodata & \nodata & Non Eruptive \\	
SP & 20070116 & 16:26 & 17:56 & 16:47 &  17:56 & Multiple jets \\
SP & 20070116 & 16:54 & 17:20 & 16:33 &  16:51 &  \\
SP & 20070117 & 12:12 & 12:45 & 12:13 &  12:46 &  \\
SP & 20070117 & 12:08 & 12:28 & \nodata & \nodata & Non Eruptive   \\
SP & 20070120 & 12:32 & 12:33 & 12:34 & 12:52 &   \\
SP & 20070120 & 15:01 &  15:09 & 15:06 & 150735 & \\

SP & 20070407 & 00:53 & \nodata & 03:31 & 03:4338 & very tiny \\
SP & 20070407 & 05:35 &  06:25 & \nodata & \nodata & \\ 	
SP & 20070407 & 18:27 &  18:55 & \nodata & \nodata & Non Eruptive \\ 

SP & 20070408 & 17:22 &  17:30 & \nodata & \nodata & Non Eruptive \\ 
SP & 20070408 & 19:06 &  19:15 & 19:12 & 1928 & \\
SP & 20070408 & 21:06 & \nodata & 21:06 & \nodata &  \\
SP & 20070408 & 22:48 & 22:56 & 22:48 & 22:59 &  \\
SP & 20070408 & 22:45 & \nodata &  23:31 & 23:35 & \\

NP & 20070910 & 20:40 & 21:07 & 20:40 & 21:18 & Multiple jets\\

NP & 20070911 & 18:22 & 18:26 & 18:20 & 18:34 & \\
NP & 20070911 & 20:02 & 20:12 & 20:04 & 20:30 & \\
NP & 20070911 & 20:08 & 20:10 & \nodata & \nodata & Non Eruptive\\
NP & 20070911 & 21:11 & 21:52 & 21:15 & 21:52 & \\

NP & 20070913 & 15:32 & 16:19 & 16:27 & 16:57 & \\
NP & 20070913 & 15:46 & 16:09 & 18:34 & 18:43 & Sigmoid reappeared during jet\\
NP & 20070913 & 17:30 & 17:58 & 17:34 & 17:46 & Quiet Sun\\

NP & 20071031 & 01:20 & 01:32 & 01:38 & \nodata & \\
NP & 20071031 & 01:00 & 03:00 & \nodata & \nodata & Non Eruptive \\	

NP & 20071101 & 03:13 & 03:27 & 03:26 & 03:45 & \\
NP & 20071101 & 23:59 & 00:04 & 00:20 & \nodata & \\

NP & 20071103 & 11:17 & 11:21  & 11:17  & 11:27 & \\
NP & 20071103 & 11:58 & 12:18  & 11:52  & 12:22 &  \\
NP & 20071103 & 12:36 & 12:39  & 12:37  & 12:47 & \\

NP & 20071104 & 06:32 & 06:33  & 06:29  & 06:38 & \\

\enddata
\end{deluxetable}

\end{document}